\documentclass[twocolumn,prd,longbibliography]{revtex4}
\usepackage{graphicx}

\newcommand{\R}{{\bf R}}

\newcommand{\CO}{{\cal O}}

\newcommand{\bk}{{\bf k}}

\newcommand{\by}{{\bf y}}

\newcommand{\p}{\partial}

\renewcommand{\bar}[1]{\overline{#1}}
\renewcommand{\tilde}[1]{\widetilde{#1}}
\newcommand{\be}{\begin{equation}}
\newcommand{\ee}{\end{equation}}
\newcommand{\bea}{\begin{eqnarray}}
\newcommand{\eea}{\end{eqnarray}}
\usepackage{graphicx}

\begin{document}
\title{Nonrelativistic Short-Distance Completions of a Naturally Light Higgs}
\author{Kevin T. Grosvenor${}^{a}$, Petr Ho\v{r}ava${}^{b,c}$, 
Christopher J. Mogni${}^{b,c}$ and Ziqi Yan${}^{b,c}$}
\affiliation{\smallskip
${}^a$Niels Bohr Institute, Copenhagen University\\ 
Blegdamsvej 17, DK-2100 Copenhagen \O , Denmark\smallskip\\
${}^b$Berkeley Center for Theoretical Physics and Department of Physics\\ 
University of California, Berkeley, California 94720-7300\smallskip\\
${}^c$Physics Division, Lawrence Berkeley National Laboratory\\ 
Berkeley, California 94720-8162}
\begin{abstract} 
Nonrelativistic scalar field theories can exhibit a natural cascading hierarchy of scales, protected by a hierarchy of polynomial shift symmetries.  Using a simple model, we argue that a high-energy cross-over to such nonrelativistic behavior naturally leads to light scalars, and thus represents a useful ingredient for technically natural resolutions of scalar mass hierarchies, perhaps even the Higgs mass hierarchy puzzle.  
\end{abstract}
\maketitle
\section{Introduction}

Two of the most prominent puzzles of fundamental physics -- the cosmological constant problem and the Higgs mass hierarchy problem -- can be viewed as puzzles of technical naturalness \cite{th}.  The 2012 discovery \cite{atlas,cms} and the observed properties of the Higgs boson suggest that the Standard Model may be self-contained up to a very high scale.  This intriguing possibility brings the naturalness puzzles back into renewed focus (see {\it e.g.} \cite{gian,gian8}), and invites us to look for new ideas about naturalness.

There are two perspectives on naturalness: Technical naturalness, formalized by 't~Hooft \cite{th}, states that a parameter may be naturally small if setting it to zero leads to an enhanced symmetry.  The stronger naturalness in the sense of Dirac \cite{dirac,diracc} simply requires that there be no ``unexplained'' small numbers in Nature.  Our philosophy in this paper is to search for mechanisms which produce {\it technical\/} naturalness: explaining the permissibility of small numbers, but not necessarily their origin.  

In the past few years, we have learned that the concept of technical naturalness exhibits many surprises in nonrelativistic settings (see \cite{sur} for a brief  review).  New symmetries emerge \cite{msb}, and they protect new hierarchies of Nambu-Goldstone bosons with cascading scales of partial symmetry breaking \cite{cmu}. In this paper, we apply this phenomenon to relativistic scalars such as the Higgs, and investigate the possibility of a crossover to nonrelativistic physics at high energy scales and its influence on the naturalness of a small Higgs mass.

\subsection{Nonrelativistic physics and the Standard Model}

For gravity, nonrelativistic physics with possible fundamental anisotropies between space and time is beneficial for improving the short-distance behavior, possibly leading to a UV complete theory \cite{lif,mqc}.  In contrast, there seem to be no similar benefits from viewing the Standard Model (SM) as a low-energy effective description of an underlying fundamentally nonrelativistic theory:  The SM is already renormalizable and nearly UV complete (assuming that one can remedy the growth of the hypercharge coupling at high energies).  However, an embedding of the SM into a nonrelativistic theory may be of interest, if it provides a way out of the Higgs mass hierarchy problem without ruining the observed Lorentz symmetry at accessible energies.  

\subsection{Lightning review of the naturalness puzzle for the scalar masses}

The essence of the naturalness problem of a light scalar with nonderivative self-interactions (such as the Higgs) can be succinctly illustrated by considering a single relativistic scalar $\Phi(x^\mu)$ in $3+1$ dimensions with action  
\be
\label{actr}
S=\frac{1}{2}\int d^4x\left(\p_\mu\Phi\p^\mu\Phi-m^2\Phi^2-\frac{\lambda}{12}\Phi^4\right).
\ee
Can $m^2$ be small?  Not independently of the value of $\lambda$: both nonderivative terms in (\ref{actr}) break the same, constant shift symmetry $\Phi\to\Phi+\delta\Phi$ with $\delta\Phi=b$, and therefore must be of the same order of smallness (measured by $\varepsilon\ll 1$) relative to the naturalness scale $M$: 
\be
\label{natrel}
m^2\sim\varepsilon M^2,\qquad\lambda\sim\varepsilon.
\ee
This gives the following simple but important relation, 
\be
\label{natrelsc}
M\sim\frac{m}{\sqrt{\lambda}},
\ee
which then implies the naturalness problem:  $m$ cannot be made arbitrarily smaller than $M$ without $\lambda$ being made correspondingly small to assure that the naturalness condition (\ref{natrelsc}) hold.  At typical values of $\lambda$ not much smaller than 1, $m$ will be of the order of the naturalness scale $M$, ruining the hierarchy.  Note that this naturalness problem is present already before gauging. 

Finding new ways around relations (\ref{natrel}) without putting technical naturalness in jeopardy is the main goal of this paper.

\section{A toy model}

For simplicity, and to highlight the novelties associated with nonrelativistic naturalness, let's consider a simple toy model first:  The theory of a real scalar field $\phi(t,\by)$ in $3+1$ dimensions, $\by=(y^i,i=1,\ldots, 3)$, with Aristotelian spacetime symmetry.  The terminology is that of \cite{penb} (see \cite{sur}): The {\it Aristotelian spacetime\/} is defined as $\R^{3+1}$ with the flat metric and the preferred foliation by constant time slices; and the {\it Aristotelian symmetries\/} contain spatial rotations and translations and time translations, but no boosts (neither Lorentzian nor Galilean).  Such spacetimes emerge naturally in the context of nonrelativistic gravity \cite{lif,mqc}, as the ground-state solutions of the theory with zero cosmological constant.

Our model was introduced in \cite{cmu} and studied further in \cite{nrr}.  We start with the free theory controlled at short distances by the Gaussian fixed point with dynamical exponent $z=3$ ($z$ is the measure of the anisotropy between time and space).  The action is
\bea
\label{stwo}
S_2&=&\frac{1}{2}\int dt\,d^3\by\left\{\dot\phi^2-\zeta_3^2\p_i\p_j\p_k\phi\,
\p_i\p_j\p_k\phi\right.\cr
&&\qquad\left.{}-\zeta_2^2\p_i\p_j\phi\,\p_i\p_j\phi-c^2\p_i\phi\,\p_i\phi
-m^2\phi^2\vphantom{\dot\phi}\right\}.
\eea
The first two terms define the Gaussian $z=3$ fixed point, and the remaining three terms are its relevant Gaussian deformations.  Classically, we can set $\zeta_3^2=1$ by a one-time rescaling of space-time coordinates.  We measure the dimensions in the units of energy; with the scaling of the $z=3$ fixed point, we have:  $[t]=-1$, $[y^i]=-1/3$, $[\p_i]=1/3$, and the field $\phi$ is dimensionless (i.e., at its {\it lower critical dimension\/}).  The dimensions of the relevant Gaussian couplings are $[\zeta_2^2]=2/3$, $[c^2]=4/3$ and $[m^2]=2$.

To this free action, we add interaction terms $S_{\rm int}$ whose choice depends on the desired symmetries; in the simplest model, we take
\be
\label{sinto}
S_{\rm int}=-\frac{\lambda_3}{2}\int dt\,d^3\by\,\CO
-\frac{\lambda_0}{4!}\int dt\,d^3\by\,\phi^4,
\ee
where 
\be
\label{ooo}
\CO=\p_i\phi\,\p_i\p_j\phi\,\p_j\p_k\phi\,\p_k\phi+\frac{1}{3}\p_i\phi\,\p_j\phi\,\p_k\phi\,\p_i\p_j\p_k\phi.
\ee
The four-point six-derivative self-coupling constant $\lambda_3$ is marginal ($[\lambda_3]=0$).  With Higgs applications in mind, we also added the nonderivative $\phi^4$ self-interaction.  In our microscopic theory, its coupling is relevant, $[\lambda_0]=2$.

\subsection{Symmetries}

Scalar field theories on the Aristotelian spacetimes can exhibit {\it polynomial shift symmetries\/} \cite{msb,pol} of degree $P$, generated by 
\be
\delta\phi(t,\by)=b+b_iy^i+b_{ij}y^iy^j+\ldots+b_{i_1\ldots i_P}y^{i_1}\cdots y^{i_P},
\ee
with $b_{i_1\ldots}$ real constants.  
$P$ can be any non-negative integer, but we will only be interested in $P\leq 2$.  Our theory with the interaction term given by (\ref{ooo}) is power-counting renormalizable, since at the $z=3$ fixed point $\CO$ is {\it the only marginal or relevant interaction term invariant under the linear shift symmetry}.  (With the symmetry reduced to the constant shifts, there would also be one relevant and three additional marginal interaction terms.)

\subsection{Quantum properties and renormalization}

The theory with the linear-shift invariant interaction given by the $\lambda_3$ term in (\ref{sinto}) (and with $\lambda_0=0$) exhibits intriguing quantum properties \cite{cmu,nrr}.  There is no wave-function renormalization of $\phi$, and both $\lambda_3$ and $c^2$ satisfy a non-renormalization theorem to all orders in $\lambda_3$.  This does not mean that the theory can be weakly coupled at all scales:  The coefficient $\zeta_3^2$ in (\ref{stwo}) -- which we set equal to 1 in the classical limit -- is logarithmically divergent starting at two loops, and therefore $\zeta_3$ runs with the renormalization-group (RG) scale.  The effective coupling in the two-on-two scattering amplitude is not $\lambda_3$ but $\bar\lambda\equiv\lambda_3/\zeta_3^3$, which runs due to the running of $\zeta_3$.  The two-loop $\beta$ function reveals that $\bar\lambda$ increases with increasing energy.  

The theory exhibits interesting instabilities.  First, note that the contribution of the $\lambda_3$ term in (\ref{sinto}) to the Hamiltonian is unbounded both below and above.  Assuming that the couplings in (\ref{stwo}) have been chosen such that the dispersion relation is positive definite, the theory is perturbatively stable around $\langle\phi\rangle=0$.  However, at nonzero $\lambda_3$, this state is non-perturbatively unstable, with the decay probability controlled by a bounce instanton \cite{sc,sccc}.  The analytic form of the dominant instanton is not known, but its contribution to the decay rate of the vacuum will be $\Gamma_{\rm vac}\propto\exp(-C/\bar\lambda)$, with $C$ positive and of order one (assuming at least one of the infrared-regulating couplings $m^2$, $c^2$ or $\zeta_2^2$ is non-zero).  If one so desires, this vacuum instability can be cured by embedding this model into a (more complex but stable) theory with the symmetries reduced to constant shifts \cite{nrr}.  

The $\phi$ quanta also exhibit an intriguing perturbative instability in the Aristotelian spacetime:  A single particle with a large enough momentum $\bk$ with respect to the rest frame acquires a non-zero decay width $\Gamma$.  This is the usual phenomenon of quasiparticle damping known from condensed matter.  The leading contribution to $\Gamma$ arises at two loops, and at large $\bk$ goes as $\Gamma\sim\zeta_3\bar\lambda{}^2|\bk|^3$.  The $\phi$ quantum is absolutely stable at $\bk\approx 0$; at small $|\bk|$ above the threshold, we find a very sharp long-lived resonance. With increasing $|\bk|$, $\bar\lambda$ runs until it becomes $\sim 1$ and the $\phi$ resonance becomes very wide.  The scattering of individual quanta with very high $|\bk|$ makes no sense: They decay rapidly into multiple soft quanta beforehand.  Thus, one cannot simply argue that the running of the $\bar\lambda$ coupling to large values makes the theory UV incomplete:  These large values of $\bar\lambda$ cannot be measured by any two-on-two scattering; only softer quanta are available to scatter and the theory may self-complete in a mechanism reminiscent of classicalization \cite{gia1,gia0}.  In this paper, we do not need any such completion and will use this model only up to a very high naturalness scale where the coupling will still be small. 

\section{Nonrelativistic vs relativistic observers and Naturalness}

While the theory is unambiguously defined by its microscopic behavior around the $z=3$ fixed point, its physical properties will look somewhat different to different observers.  

\subsection{Nonrelativistic observers and the microscopic theory}

Unlike in relativistic theories, there are at least two natural classes of observers already at the microscopic level.  The {\it Aristotelian observers\/} fix the coordinates $(t,\by)$ once and for all, find the non-renormalization of $\lambda_3$ but also the running of $\zeta^2_3$, which lead to the running of the effective coupling $\bar\lambda$.  The {\it Wilsonian observers}, during the process of integrating out a shell of modes, rescale the system to restore the normalization condition $\zeta_3^2=1$.  This involves a rescaling of the spatial coordinates which depends on the RG scale.  Their effective coupling runs.  When they compare their notes with the Aristotelian observers, both see the same physics but in a slightly rescaled coordinate system.  

\subsection{Effective relativistic observers at low energies}

At low energies, the lowest-derivative terms dominate.  The higher-derivative terms are suppressed, and the system develops an accidental approximate Lorentz symmetry, with small Lorentz-violating corrections.  Observers at those energies will find it natural to interpret the system relativistically.  We shall refer to such observers as ``low-energy relativistic observers.''  While for the microscopic observers $c^2$ is a relevant coupling, for the low-energy relativistic observers $c$ appears to be a constant of nature, insofar as they cannot detect deviations from the constancy of $c$ due to the small Lorentz-violating terms.  Given their relativistic prejudice, their natural coordinate frame is 
\be
x^0=t,\qquad x^i=y^i/c.  
\ee
Note that this gives the correct dimensions of a relativistic coordinate system, $[x^0]=[x^i]=-1$.  In these coordinates, the low-energy relativistic observer finds the action of our system to be
\bea
\label{actlow}
S&=&\frac{1}{2}\int d^4x\left\{\nabla_\mu\Phi\nabla^\mu\Phi-m^2\Phi-\frac{1}{12}\lambda_h\Phi^4\right.\cr
&&\left.{}-\tilde\zeta_3^2(\nabla_i\nabla_j\nabla_k\Phi)^2-\tilde\zeta_2^2(\nabla_i\nabla_j\Phi)^2-\tilde\lambda_3\tilde\CO\vphantom{\frac{1}{12}}\right\}
\eea
where $\nabla_\mu\equiv\p/\p x^\mu$, $\Phi=c^{3/2}\phi$ is the scalar field properly rescaled to match the perspective of the relativistic observer, and $\tilde\CO$ is given by (\ref{ooo}) with $\p$ replaced by $\nabla$ and $\phi$ by $\Phi$.  
The low-energy parameters are given in terms of the microscopic parameters as follows.  For the relativistic observer, the mass $m$ of $\Phi$ is equal to the gap parameter $m$ of the microscopic theory, and its nonderivative self-coupling is given by
\be
\label{higgsmic}
\lambda_h=\lambda_0/c^3.
\ee
The remaining couplings in (\ref{actlow}) are given in terms of the microscopic parameters by
\be
\label{nonrcor}
\tilde\zeta_3^2=\zeta_3^2/c^6,\quad\tilde\zeta_2^2=\zeta_2^2/c^4,\quad 
\tilde\lambda_3=\lambda_3/c^9;
\ee
from the low-energy perspective, they represent irrelevant terms which violate Lorentz invariance.  

\subsection{Microscopic naturalness}

In accord with the principles of causality, we require that technical naturalness hold at the level of the microscopic, nonrelativistic theory.  Technically natural hierarchies with varying degrees of complexity are possible \cite{cmu}.  In the simplest, one $\varepsilon$ controls the breaking of the linear shift symmetry to no shift symmetry at all, and all couplings are of order $\varepsilon$ in units of the naturalness momentum scale $\mu$.  However, this crude pattern may be naturally refined:  $\lambda_3$ and $\zeta_2^2$ preserve linear shifts and can be controlled by their own smallness parameter: $\lambda_3\sim\varepsilon_2$ and $\zeta_2^2\sim\varepsilon_2\mu^2$.   At this stage, quantum corrections to $c^2$ will not be generated (due to the nonrenormalization theorem), and can be kept of order $\varepsilon_1\ll\varepsilon_2$.  Finally, the nonderivative terms break constant shift symmetry, and can be of order $\varepsilon_0\ll\varepsilon_1$.  We thus obtain a technically natural cascading hierarchy of scales.  This cascade is associated with a natural hierarchy of crossover scales:  At very high scales (around $\mu$), the system is dominated by the $z=3$ scaling, then it crosses at lower scales to a $z=2$ regime, followed by another crossover to the $z=1$ regime, until it finally reaches the lowest scales set by the gap $m$.  Next we need to examine how this cascading hierarchy appears from the viewpoint of the low-energy relativistic observer.  

\subsection{Naturalness for relativistic low-energy observers}

We begin at the microscopic level, with the following, technically natural hierarchy of couplings, 
\bea
\label{natmic}
&&\zeta_3^2\sim 1,\qquad\lambda_3\sim\varepsilon_2,\qquad\zeta_2^2\sim\varepsilon_2\mu^2,\nonumber\\
&&\quad c^2\sim\varepsilon_1\mu^4,\qquad m^2\sim\lambda_0\sim \varepsilon_0\mu^6,
\eea
and 
\be
\varepsilon_0\ll\varepsilon_1\ll\varepsilon_2\ll 1.
\ee
What are the sizes of the couplings that the low-energy relativistic observer will see?  Plugging (\ref{natmic}) into (\ref{higgsmic}) and (\ref{nonrcor}), and introducing the naturalness energy scale $M\equiv\mu^3$, we obtain
\be
\label{higgslow}
m^2\sim\varepsilon_0 M^2,\qquad \lambda_h\sim\varepsilon_0/\varepsilon_1^{3/2}
\ee
for the scalar mass and self-coupling, and 
\be
\label{lorlow}
\tilde\zeta_3^2\sim\frac{\varepsilon_0^2}{\varepsilon_1^3}\frac{1}{m^4}, \quad
\tilde\zeta_2^2\sim\frac{\varepsilon_0\varepsilon_2}{\varepsilon_1^2}\frac{1}{m^2}, \quad
\tilde\lambda_3\sim\frac{\varepsilon_0^3\varepsilon_2}{\varepsilon_1^{9/2}}\frac{1}{m^6}
\ee
for the irrelevant nonrelativistic corrections.  

This is the central result of this paper:  In contrast to the standard relativistic relations (\ref{natrel}), we now have a new small parameter $\varepsilon_1$ which controls $c^2$, modifies the relations to (\ref{higgslow}), and makes a technically natural large hierarchy of scales with sizable values of the coupling $\lambda_h\sim 1$ possible.

\section{Towards the Higgs and the Standard Model}

Now we would like to couple this naturally light scalar to the rest of the SM.  We will assume the Higgs-less part of SM to be exactly relativistic until the coupling to the Higgs; the coupling will induce violations of Lorentz invariance that we wish to keep naturally small in order to conform to the stringent experimental bounds on Lorentz violations \cite{stefano13}, and without spoiling the mass hierarchy.  

For simplicity, we will continue working within the logical structure of our toy model, but the results are more universal, robust and model-independent.  First, as noted in \cite{nrr}, our toy model can be extended to a unique theory with global $SO(N)$ symmetry with $\phi$ in the $N$.  The case of $N=4$ will correspond to the candidate Higgs; with the flip of the $m^2$ sign, $\phi$ will develop a condensate $\langle\phi\rangle=m/\sqrt{\lambda_0}\sim 1$.

\subsection{Hierarchy between $M_{EW}$ and $M_X$}

Phenomenologically, we would like $m\sim M_{EW}$ of the order of the electroweak scale, while $M\sim M_X$ of the order of some high scale $M_X$, such as the Planck scale or a GUT scale.  To illustrate our mechanism, we will try to go the whole hog and realize a hierarchy across 15 orders of magnitude, between the electroweak scale and the Planck scale.  For numerical simplicity, we take $m\sim 1$ TeV and $M\sim 10^{18}$ GeV.  At the same time, we want the Higgs self-coupling $\lambda_h$ not too much smaller than $\sim 1$.

As a very simple and concrete example, take the following ``10-20-30'' model:
\be
\label{123}
\varepsilon_2\sim 10^{-10},\qquad 
\varepsilon_1\sim 10^{-20},\qquad 
\varepsilon_0\sim 10^{-30}.
\ee
From (\ref{higgslow}), we obtain
\be
m/M\sim 10^{-15},\quad \lambda_h\sim 1,
\ee
precisely as desired!  (Smaller $\lambda_h$, say $\sim 0.1$, are easily arranged by small changes of (\ref{123}).)  Moreover, the irrelevant Lorentz-violating couplings (\ref{lorlow}) are pushed above the TeV scale:
\be
\tilde\zeta_3^2\sim\frac{1}{m^4},\quad \tilde\zeta_2^2\sim\frac{1}{m^2},\quad
\tilde\lambda_3\sim10^{-10}\frac{1}{m^6}.
\ee
The $\tilde\zeta^2$ couplings yield small nonrelativistic modifications of the Higgs dispersion relation $\omega^2=m^2+\bk^2$ by higher power terms $|\bk|^4$ and $|\bk|^6$, representing the first observable signatures of the ``new physics'' that cures the hierarchy problem:  the Higgs sector exhibits a crossover towards $z>1$ at scales of order $m\sim 1$ TeV.  Pushing the nonrelativistic corrections to higher scales should be possible in slightly more sophisticated versions of our simplest 10-20-30 scenario; for example, making $\lambda_h<1$ further suppresses the size of $\tilde\zeta_3^2$, since $\tilde\zeta_3^2\sim\lambda_h^2/m^4$; and $\tilde\zeta_2^2$ can be suppressed by simply choosing a smaller $\varepsilon_2$.

\subsection{Fermions and Yukawa couplings}

We can couple the scalar $\phi$ to several species of relativistic fermions $\Psi_f(t,\by)$, whose two chiralities we assume to be in distinct representations to prevent bare masses (as in the SM).  In the microscopic theory, their dimension is $[\Psi_f]=1/2$.  Their relativistic kinetic term written in nonrelativistic coordinates is
\be
\sum_f\int dt\,d^3\by\,(\Psi_f^\dagger\dot\Psi_f+c_f\bar\Psi_f\gamma^i\p_i\Psi_f).
\ee
Before coupling to $\phi$, all fermions see the same limiting speed, which we set equal to $c_f=c$.  When we couple the fermions to $\phi$, their dispersion relation acquires nonrelativistic corrections from Higgs loops; we need these to be small, without spoiling the Higgs mass hierarchy.  The most relevant coupling of $\Psi_f$ to $\phi$ is the Yukawa term
\be
\sum_f Y_f\int dt\,d^3\by\,\bar\Psi_f\phi\Psi_f.
\ee
When non-zero, the Yukawa couplings $Y_f$ break the constant shift symmetry of $\phi$, and one may expect them all to be bounded from above by the parameter $\varepsilon_0$ which controls all the other terms breaking the constant shift symmetry in the Higgs sector.  (This is in the units of $\mu^3$, since $[Y_f]=1$.)  However, there is some wiggling room:  Detailed estimates of the Higgs loop corrections show that we can increase the range of the Yukawas to include the window from $\varepsilon_0$ to $\sqrt{\varepsilon_0}$, without spoiling the smallness of $m^2$ and $\lambda_0$.  This requires that we also include the nonrelativistic terms $\zeta_{3f}\bar\Psi\gamma^i\p_i\p^2\Psi$ with $\zeta_{3f}\lesssim 1$ to the action (which will be generated by the Higgs loops anyway).  This increase in the range of the Yukawas works because the corrections to $m^2$ and $\lambda_0$ due to fermionic loops are at least quadratic in $Y_f$'s.  Thus, the window of naturalness for the non-zero Yukawas has been extended to include  
$\varepsilon_0\mu^3\lesssim Y_f\lesssim\sqrt{\varepsilon_0}\mu^3,$ 
self-consistently requiring that $\zeta_{3f}\sim Y_f^2/m^2$ for each fermion.  

The low-energy relativistic observer rewrites the theory in terms of the naturally normalized fermions $\psi_f(x^\mu)=c^{3/2}\Psi_f(t,\by)$, and sees the Yukawa terms as 
\be
\sum_f y_f\int d^4 x\,\Phi\bar\psi_f\psi_f,
\ee
with $y_f=Y_f/c^{3/2}$.  The naturalness window for the Yukawas as seen by the relativistic observer extends to 
\be
\label{extyu}
y_f\lesssim\varepsilon_0^{1/2}/\varepsilon_1^{3/4}.  
\ee

This extension past the naive bound $y_f\lesssim\varepsilon_0/\varepsilon_1^{3/4}$ is crucial:  In our 10-20-30 scenario, the naive bound would require $y_f\lesssim 10^{-15}$, excluding all fermions.  The extended bound (\ref{extyu}) requires $y_f\lesssim 1$, a range which naturally accommodates the masses of all the known fermions, from the top quark at the upper bound, down to the likely values of the neutrino masses not too far above the naive bound.  Thus, in the 10-20-30 scenario, all the SM fermion masses can be Dirac masses, in a technically natural way.

\subsection{Gauging}

Next we couple the system to relativistic Yang-Mills fields.  In the microscopic theory, this is done by covariantizing the derivatives to $D_t\phi\equiv\dot\phi+iea_0\phi$ and $D_i\phi=(\p_i+iea_i)\phi$.  We normalize the gauge fields such that when rewritten in the nonrelativistic language, their standard relativistic action is $\int dt\,d^3\by\,\{\frac{1}{2}(\p_ia_0-\dot a_i+\ldots)^2-(c^2/4)(\p_ia_j-\p_ja_i+\ldots)^2\}$.  Thus, we have $[a_i]=0$, $[a_0]=2/3$, and the gauge coupling is relevant, $[e]=1/3$.  The low-energy relativistic fields $A_\mu$ and the Yang-Mills coupling are related to these microscopic variables by $A_i=c^{3/2}a_i$, $A_0=c^{1/2}a_0$, and $g_{\rm YM}^2=e^2/c$.

What is the size of the Yang-Mills coupling $g_{\rm YM}$ seen by the low-energy observer?  The microscopic gauge coupling $e$ breaks the constant shift symmetry of $\phi$.  Hence, the gauge loops can be expected to correct $m^2$ by $\sim e^2\mu^4$.  To maintain naturalness, this would require $e^2\sim\varepsilon_0\mu^2$; the low-energy observer would then find the Yang-Mills coupling $g_{\rm YM}=e/c^{1/2}\sim\varepsilon_0^{1/2}/\varepsilon_1^{1/4}$.  Unfortunately, if these estimates are accurate ({\it i.e.}, in the absence of additional cancellations or hidden symmetries), it would be very difficult to make $g_{\rm YM}\sim 1$ while keeping $\lambda_h\sim 1$.  In particular, in our simple 10-20-30 model the natural values of the gauge couplings come out unrealistically small, $g_{\rm YM}\sim 10^{-10}$, implying unrealistically light gauge bosons.  It is at present unclear whether these estimates can be improved to achieve a scenario with more realistic values of $g_{\rm YM}$; this question would require a more detailed analysis of the interplay between polynomial shift symmetries and gauge symmetries, beyond the scope of this paper.

\subsection{$z=2$ or $z=3$?}

How important is it to embed the Higgs into a $z=3$ theory?  Can we choose the simpler $z=2$ short-distance behavior, perhaps improving the prospects of a realistic gauging?  Interestingly, the answer is no, if we insist on $\lambda_h$ in (\ref{higgslow}) to be ${}\sim 1$:  In the absence of $z=3$ terms, the leading nonrelativistic corrections originate from $\zeta_2^2\sim 1$, and they become important at unacceptably low energies $\ll m$.  The $z=2$ observer would be almost as mystified about the naturalness of a light Higgs as the relativistic observer.  Thus, $z=3$ is the lowest value of $z$ in the microscopic system for which our mechanism with relations (\ref{higgslow}) and $\lambda_h\sim 1$ can work, without generating large Lorentz violations at low energies.  

\section{Conclusions}

In this paper, we have presented a new mechanism leading to naturally light scalars whose nonderivative self-couplings can be large.  The mechanism involves a crossover from the low-energy relativistic regime to a highly nonrelativistic high-energy regime.  In the underlying nonrelativistic theory, the effective low-energy speed of light $c^2$ can be naturally small, allowing naturally light scalars with masses $m\ll M$ much smaller than the naturalness scale $M$.  When coupled to fermions, this mechanism yields an appealing structure of Yukawa couplings.  However, while the construction can be extended to include gauge fields with the scalar interpreted as a Higgs, our simplest 10-20-30 scenario does not quite work for the SM Higgs: the gauge couplings and consequently the $W$ and $Z$ boson masses are simply too small.  

Our results have been based on rather conservative estimates of the quantum corrections, ensuring but not necessarily optimizing naturalness.  These estimates can certainly be further tightened, refined by invoking more symmetries, or otherwise improved.  In particular, it should be noted that we have not relied on (nor included) the omnipresent loop suppression factors involving powers of $\sim 1/(16\pi^2)$.  A more detailed investigation is needed before we can conclude whether our mechanism is a useful ingredient for resolving the Higgs mass hierarchy puzzle in the SM\@.  It is clear, however, that our results about naturalness are relevant to other scalar fields, with or without gauge invariance, including the inflaton.  

{\bf Acknowledgements:} This work has been supported by NSF Grant PHY-1521446 and by Berkeley Center for Theoretical Physics.  One of us (P.H.) has been supported in part by the Sabbatical Fellowship Grant \#399747 from the Simons Foundation; one of us (K.T.G.) has been supported by ERC Advanced Grant 291092 ``Exploring the Quantum Universe''; and one of us (C.J.M.) has been supported by a Graduate Research Fellowship from NSF GRFP.
\bibliography{res}
\end{document}